# Magnetism in $MoS_2$ induced by MeV proton irradiation

S. Mathew,[1,2,a)] K. Gopinadhan,[1,2] T. K. Chan,[3,6] X. J. Yu,[2,4] D. Zhan,[5] L. Cao,[6] A. Rusydi,[2,4,6] M. B. H. Breese,[3,4,6] S. Dhar,[1,2] Z. X. Shen,[5] T. Venkatesan,[1,2,b)] and John TL Thong,[1,c)]

[1]*Department of Electrical and Computer Engineering, National University of Singapore, Singapore, 117576*
[2]*NUSNNI-NanoCore, National University of Singapore 117576*
[3]*Center for Ion Beam Applications (CIBA), Department of Physics, National University of Singapore, Singapore 117542*
[4]*Singapore Synchrotron Light Source (SSLS), National University of Singapore, Singapore 117603*
[5]*Division of Physics and Applied Physics, School of Physical and Mathematical Sciences, Nanyang Technological University, Singapore 637371*
[6]*Department of Physics, National University of Singapore, Singapore 117542*

*Molybdenum disulphide, a diamagnetic layered dichalcogenide solid, is found to show magnetic ordering at room temperature when exposed to a 2 MeV proton beam. The temperature dependence of magnetization displays ferrimagnetic behavior with a Curie temperature of 895 K. A disorder mode corresponding to a zone-edge phonon and a Mo valence higher than +4, have been detected in the irradiated samples using Raman and X-ray photoelectron spectroscopy, respectively. The possible origins of long-range magnetic ordering in irradiated $MoS_2$ samples are discussed.*

Authors to whom correspondence should be addressed.
a) pmsmathew@gmail.com, b) venky@nus.edu.sg and c) elettl@nus.edu.sg

## 1. Introduction

Magnetic ordering in materials that do not have partially filled *d* or *f* orbitals is a topic of interest in recent years. [1-5] MoS$_2$ is one of the central members in transition-metal-dichalcogenide compounds. [6] Its layered structure, held together with van der Waals forces between the layers, along with its remarkable electronic properties such as charge density wave transitions in transition-metal-dichalcogenides make the material interesting from both fundamental and applied research perspectives. [6-8] Furthermore, monolayer MoS$_2$ is a semiconducting analogue of graphene and has been fabricated recently. [9,10]

An interesting weak ferromagnetism phenomenon in nano-sheets of MoS$_2$ had previously been reported by Zhang et al. who attributed the observed magnetic signal to the presence of unsaturated edge atoms. [11] There have been several theoretical efforts in understanding ferromagnetic ordering in MoS$_2$. Using density functional theory, Li et al. predicted ferromagnetism in zigzag nanoribbons of MoS$_2$. [12] Formation of magnetic moments was also reported in Mo$_n$S$_{2n}$ clusters, [13] nanoparticles[14] and nanoribbons[15,16] of MoS$_2$ from first principle studies. However, to date, there have been no experimental observations of ferromagnetism in modified bulk crystals of MoS$_2$.

Ion irradiation introduces a wide range of defects in a controlled manner and is used to tailor the structural and even the magnetic properties of materials. Ferromagnetic ordering has been realized in MeV proton irradiated highly-oriented pyrolitic graphite (HOPG)[17] and fullerene thin films. [18] Very recently, Makarova et al. reported magnetism in keV H$^+$ and He$^+$ irradiated HOPG.[19] Furthermore, Ohldag et al. showed that magnetism in proton-irradiated graphite originates only from the carbon π-electron system using X-ray magnetic circular dichroism.[20] The discovery of ferromagnetism in $MoS_2$ nano-sheets along with various simulation studies and the above reports on magnetism in carbon allotropes raise the possibility of magnetism in ion- irradiated $MoS_2$ system.

Here we present results of magnetization measurements on pristine and 2 MeV proton irradiated $MoS_2$ samples. We find that magnetism is induced in proton-irradiated $MoS_2$ samples. The observation of long-range magnetic ordering in ion-irradiated $MoS_2$ points towards the possibility of selectively fabricating magnetic regions in a diamagnetic matrix, which may enable the design of unique spintronic devices.

## 2. Experimental.

Samples for irradiation were prepared in the following manner. $MoS_2$ flakes, of 2 mm diameter and ~200 μm in thickness, were glued with diamagnetic varnish onto a high-purity silicon substrate. Ion irradiations were carried out at room temperature using the 3.5 MV Singletron facility at the Center for Ion Beam Applications at the National University of Singapore. Magnetic measurements were performed using superconducting quantum interference device system (MPMS SQUID-VSM) with a sensitivity of $8 \times 10^{-8}$ emu. Raman spectroscopy, proton-induced X-ray emission (PIXE) and X-ray photoelectron spectroscopy (XPS) at the SINS beam line at the Singapore Synchrotron Light Source[21] were used to characterize the samples. The details of the ion irradiation and characterization are given in the supplemental material.[22]

## 3. Results and Discussions.

Results of magnetization versus field (M-H) measurements at 300 K and 10 K for the sample irradiated at a fluence of $1 \times 10^{18}$ ions/cm$^2$ and the pristine sample are shown in Fig. 1. The pristine sample is diamagnetic in nature. The appearance of hysteresis along with a clear remanence and coercivity (~700 Oe at 10 K) and its decrease with increasing temperature in the irradiated sample (Fig. 1 (a) and (b)) clearly indicate that MeV proton irradiation has induced ferro- or ferri-magnetic ordering in the MoS$_2$. The observed magnetic ordering can be due to the presence of defects such as atomic vacancies, displacements, and saturation of a vacancy by the implanted protons. The same sample was subsequently irradiated at cumulative fluences of $2 \times 10^{18}$ and $5 \times 10^{18}$ ions/cm$^2$ to probe the evolution of induced magnetism with ion fluence.

Magnetizations as a function of field isotherms at a fluence of $5 \times 10^{18}$ ions/cm$^2$ are shown in Fig. 2(a). An enlarged view of the M-H curves near the origin is given in the inset and the decrease of coercivity with increasing temperature is clear from the plot. A plot of coercivity *vs* ion fluence at various temperatures is given in Fig. 2(b). The value of coercivity is found to increase with ion fluence at all temperatures used in this study, although at 10 K it is almost constant. A large variation of coercivity with temperature indicates the presence of long-range magnetic ordering in the irradiated samples. The Zero Field Cooled (ZFC) and Field Cooled (FC) magnetizations at an applied field of 500 Oe are given in Fig. 2(c). An insight into the nature of magnetic ordering (ferro- or ferri-magnetic) can be gained by analyzing the variation of susceptibility with temperature. The inverse of susceptibility plot is shown as an inset of Fig. 2(c). The value of Curie temperature ($T_c$) is estimated to be ~895 K. The nature of the curve near $T_c$, a concave curvature with respect to the temperature axis, is characteristic of ferrimagnetic ordering whereas for a ferromagnetic material this curvature near $T_c$ would be convex. [23] The variation of magnetization at 300 K with an applied field of 5 kOe is plotted in Fig 2(d) for different fluences. The magnetization of the pristine sample and sample irradiated at a fluence of $1 \times 10^{17}$ ions/cm$^2$ are negative, at a fluence of $1 \times 10^{18}$ ions/cm$^2$ the sample magnetization becomes positive and at $2 \times 10^{18}$ ions/cm$^2$ the magnetism in the sample increases further, and at a fluence of $5 \times 10^{18}$ ions/cm$^2$ it has decreased. The dependence of magnetization on the irradiation fluence observed in Fig. 2(d) (the bell shaped curve) indicates that the role of the implanted protons and thus the effect of end -of-range defects for the observed magnetism is minimal, as it had been shown in the case of proton-irradiated HOPG. [19,20,24] It was demonstrated that 80% of the measured magnetic signal in the 2 MeV H$^+$ irradiated HOPG originates from the top 10 nm of the surface. [24] To probe radiation-induced modification in our samples near the surface region, we used XPS and Raman spectroscopy.

The modifications in atomic bonding and core-level electronic structure can be probed using XPS. The XPS spectra of a pristine sample and the sample irradiated at a fluence of $5 \times 10^{18}$ ions/cm$^2$ are shown in Figs. 3(a)-(d). Fitting of the spectra was done by a chi-square iteration program using a convolution of Lorentzian–Gaussian functions with a Shirley background. For fitting the Mo 3d doublet, the peak separation and the relative area ratio for 5/2 and 3/2 spin-orbit components were constrained to be 3.17 eV and 1.5, respectively, while the corresponding constraints for the S 2p 3/2 and 1/2 levels were 1.15 eV and 2, respectively.[25,26] The peaks at 228.5 eV and 231.7 eV observed in the pristine spectrum of Mo are identified as Mo $3d_{5/2}$ and $3d_{3/2}$, while the small shoulder at 226 eV in Fig. 3(a) is the sulphur 2s peak.[25] In the irradiated spectrum in Fig. 3(b), apart from the pristine Mo peaks, two additional peaks at 229.6 eV and 232.8 eV are visible. The peak observed at 229.6 eV in Fig. 3(b) has 18% intensity of the total Mo signal, which could be due to a Mo valence higher than +4. The binding energy positions of 3d levels in Mo (V) have been reported to be 2 eV higher than those of Mo (IV).[26] We found a peak at 229.6 eV in the irradiated sample which is only 1.0 eV above that of the Mo (IV) level. The pristine spectrum of S consists of S $2p_{3/2}$ and $2p_{1/2}$ peaks at 161.4 eV and 162.5 eV and another two peaks at 163 eV and 164.2 eV. The peak at 163 eV in the irradiated spectrum of S in Fig. 3(d) had increased by 6% in intensity compared to that in the pristine sample.

An indication of the nature of the induced defects and crystalline quality can be gained using Raman spectroscopy, which was a major tool for characterizing ion irradiation induced defects in graphene and graphite in a recent study.[27] The Raman spectra of the pristine and irradiated samples at a fluence of $5 \times 10^{18}$ ions/cm$^2$ are shown in Fig. 4. The $E_{2g}^{1}$ mode at 385 cm$^{-1}$ and $A_{1g}$ mode at 411 cm$^{-1}$ are clearly seen in Fig. 4.[28] In the low frequency sides of $E_{2g}^{1}$ and $A_{1g}$ phonon modes, the peaks observed in the deconvoluted spectra are the Raman-inactive $E_{1u}^{2}$ and $B_{1u}$ phonons; these modes become Raman active due to resonance effect, as observed by Sekine *et al*.[29] These extra phonons are Davydov pairs of the $E_{2g}^{1}$ and $A_{1g}$ modes.[29] The broad peak observed at ~452 cm$^{-1}$ can be the second order of LA(M) phonon.[30] In the irradiated sample, a peak at 483 cm$^{-1}$ is clearly visible. Frey et al. reported a peak at 495 cm$^{-1}$ in chemically synthesized fullerene-like and platelet-like nanoparticles of MoS$_2$ and assigned this to the second-order mode of the zone-edge phonon at 247 cm$^{-1}$. [30] Phonon dispersion and density of phonon state calculations have shown a peak ∼250 cm$^{-1}$ due to a TO branch phonon. [31] The mode at 483 cm$^{-1}$ is close to the above zone-edge phonon observed in nanoparticles of MoS$_2$. [30] The appearances of a mode at 483 cm$^{-1}$ along with the broadening of the mode at 452 cm$^{-1}$ indicate the presence of lattice defects due to proton irradiation of the sample. The FWHM of the $E_{2g}^1$ and A$_{1g}$ modes has not increased in the irradiated MoS$_2$, and this shows that the lattice structure has been preserved in the near-surface region of the irradiated sample. The A$_{1g}$ mode couples strongly to the electronic structure compared to $E_{2g}$ as observed in resonance Raman studies. [32] The ratio of the intensity of A$_{1g}$ to $E_{2g}$ modes (hereafter $R$) can be a measure of Raman cross-section as discussed in high pressure Raman spectroscopy studies. [32] The intensity ratio $R$ is found to enhance by 16% in the irradiated sample compared to the pristine sample as shown in Fig. 4. This increase in the intensity ratio $R$ in the irradiated sample can be attributed to the induced changes in electronic band structure which enhances the interaction of electrons with A$_{1g}$ phonons. [32] The intensities of the $E_{1u}^2$ and B$_{1u}$ phonon modes were also found to be enhanced in the irradiated sample. This enhancement of the Davydov pairs and second-order LA(M) peak along with the appearance of a defect mode indicate a deviation from the perfect symmetry of the system.

The magnetic moment observed for the irradiated sample at an applied field of 2000 Oe is as high as ~150 μemu along with a clear hysteresis and coercive field of 700 Oe at 10 K. The magnetization curve of the irradiated flake compared with that of the pristine sample and the blank substrate (the latter not shown here) gives clear evidence for the irradiation -induced magnetism in $MoS_2$. The total amount of magnetic impurities present in the sample was determined by post-irradiation PIXE experiments. PIXE results show ~54 ppm of Fe in the sample and if we assume all of these Fe impurities became ferromagnetic, which is a rather unrealistic assumption, the maximum signal would be only 39 μemu, far short of the observed magnetization which is at least a factor of 4 larger. Thus the observed magnetism cannot be explained by magnetic impurities. The fact that we observe diamagnetism in the pristine sample and clear magnetic hysteresis loops in the same sample after exposure to MeV protons provides irrefutable evidence for the intrinsic nature of the observed magnetic signal.

Regarding the mechanism for the formation of magnetic state in $MoS_2$, among others, the defect-mediated mechanism appears to be the most general one. Possible origins of magnetism in irradiated $MoS_2$ system could include the following: (i) irradiation-induced point defects that can give rise to magnetic moment,[16] (ii) the presence of edge states as fragments that have zigzag or armchair edges could lead to splitting of the flat energy bands and lower the energy of spin-up band compared to the spin-down band leading to ferromagnetism in the material.[11,12,14-16] Raman spectrum in Fig. 4(b) showed a defect mode at 483 $cm^{-1}$ due to the presence of zone-edge phonons in the irradiated sample. The appearance of well-defined $E_{2g}^1$ and $A_{1g}$ modes in Fig. 4(b) does not reveal any significant damage to the crystal structure of $MoS_2$ lattice. The presence of atomic vacancies, predominantly S in our case, can create a loss of symmetry and hence the coordination number of Mo atom would not remain as 6 in the irradiated sample. Tiwari et al. observed sulphur vacancy induced surface reconstruction of $MoS_2$ under high temperature treatment (above 1330 K). [33] The higher valence of Mo observed in the irradiated sample could be due to a reconstruction of the lattice, apart from the presence of sulphur vacancies. [33-35]

An estimate of the defect density created by the proton beam in $MoS_2$ can be determined from Monte Carlo simulations (SRIM 2008) using full damage cascade. [36] The displacement energy of Mo and S for the creation of a Frenkel pair used for the calculation is 20 eV and 6.9 eV, respectively, as reported by Komsa et al. in a recent study of electron irradiation hardness of transition metal dichalcogenides. [37] According to this calculation the 2 MeV proton comes to rest at a depth of 31 μm from the surface. The distance between vacancies estimated at the surface and at the end of range after irradiating with $1 \times 10^{18}$ ions/cm$^2$ is 6.3 Å and 2.7 Å, respectively. These calculated values are overestimates because the annealing of defects and crystalline nature of the target have not been incorporated in SRIM simulations.

We carried out another experiment involving low-energy proton irradiation where we subjected the $MoS_2$ sample to 0.5 MeV H$^+$ irradiation at an ion fluence of $1 \times 10^{18}$ ions/cm$^2$ (fluence at which magnetic ordering was observed using 2 MeV protons) and at a lower fluence of $2 \times 10^{17}$ ions/cm$^2$. Ferromagnetism was not observed in the former case, whereas a weak magnetic signal with a clear variation of coercivity with temperature was observed at the lower fluence in the latter case. [22] Also we have observed a weak magnetic signal with a clear hysteresis loop in the former case (the sample irradiated at a fluence of $1 \times 10^{18}$ ions/cm$^2$) after annealing (350 °C, 1 hr in Ar flow). [22] If ion damage were solely responsible for magnetization then at the end of range of the 2 MeV ion and of the 0.5 MeV ion there should be very little difference. The electronic energy loss of the ion helps recrystallization while the nuclear energy loss is only responsible for defect creation. [38,39] The ratio between electronic and nuclear energy loss is 42% greater for a 2 MeV proton compared to a 0.5 MeV proton. [36] This enhanced electronic energy loss component will allow the defective lattice to recover albeit with atomic displacements in the case of 2 MeV ion irradiation. [38,39] In the case of 0.5 MeV ions, the electronic energy loss appears to be insufficient for the required reconstruction of the lattice together with the atomic displacements to induce strong ferromagnetic signal and hence it is not effective in creating magnetic ordering in MoS$_2$.

The appearance of magnetism observed in proton-irradiated MoS$_2$ samples can be due to a combination of defect moments arising from vacancies, interstitials, deformation and partial destruction of the lattice structure, i.e., the formation of edge states and reconstructions of the lattice. For identifying the relative contributions of the ion beam induced defects towards the observed magnetism in MoS$_2$, first-principle simulations incorporating atomic vacancies, edge states and lattice-reconstructions, such as the ones performed in the case of various carbon allotropes (graphite,[40] fullerenes,[41] nano-diamond[42]) are required.

## 4. Conclusions

In conclusion, we have observed ferrimagnetic behavior in 2 MeV proton-irradiated $MoS_2$ with a Curie temperature of 895 K. Raman spectroscopy and X-ray photoelectron spectroscopy results indicate the presence of a zone-edge phonon and a Mo valence higher than +4 . Possible sources of magnetization are isolated vacancies, vacancy clusters, formation of edge states and reconstructions of the lattice. The discovery of ion-irradiation-induced magnetism in $MoS_2$ shed light on tailoring its properties using energetic ions and provides a route for future applications of this material.

## Acknowledgments.

This project is supported by an NRF-CRP grant "Graphene Related Materials and Devices". The work at SSLS is supported by NUS Core support C-380-003-002-001,

A. Rusydi and X. J. Yu also acknowledge the support from an NRF-CRP grant "Tailoring Oxide Electronics by Atomic Control," and by an MOE AcRF Tier-2 grant.

S. Mathew would like to acknowledge Ms. P. K. Ang (Dept. of Chemistry NUS, Singapore) for a set of XPS measurements and Prof. R. Nirmala (IIT Chennai, India) for fruitful discussions. S. Mathew would like to dedicate this work to the memory of Professor S. N. Behera (IOP Bhubaneswar, India) who had been an inspiration to look for magnetism in unconventional magnetic materials such as in modified carbon allotropes and other layered solids

# References

[1] T. Makarova, *Handbook of Nanophysics: Principles and Methods* (CRC press, Taylor and Francis, Oxford, 2010) p.25

[2] M. Stoneham J. Phys. Condens. Matter **22**, 74211 (2010).

[3] C. N. R. Rao, H. S. S. R. Matte, K. S. Subrahmanyam and U. Maitra, Chem. Sci. **3**, 45 (2012).

[4] A. L. Ivanovskii, Phys. Usp. **50**, 1031 (2007).

[5] I. S. Elfimov, A. Rusydi, S. I. Csiszar, Z. Hu, H. H. Hsieh, H. J. Lin, C. T. Chen, R. Liang, and G. A. Sawatzky, Phys. Rev. Lett. **98**, 137202 (2007).

[6] J. A. Wilson and A. D. Yoffe, Adv. Phys. **18**, 193 (1969).

[7] A. H. Castro Neto and K. Novoselov, Rep. Prog. Phys. **74**, 082501 (2011).

[8] H. S. S. R. Matte, A. Gomathi, A. K. Manna, D. J. Late, R. Datta, S. K. Pati, C. N. R. Rao, Angew. Chem. **122**, 4153 (2010).

[9] K. F. Mak, C. Lee, J. Hone, J. Shan, T. F. Heinz, Phys. Rev. Lett. **105**, 136805 (2010).

[10] Y. Zhan, Z. Liu, S. Najmaei, P. M. Ajayan and J. Lou, Small **8,** 966 (2012).

[11] J. Zhang, J. M. Soon, K. P. Loh, J. Yin, J. Ding, M. B. Sullivian and P. Wu, Nano Lett., **7,** 2370 (2007).

[12] Y. Li, Z. Zhou, S. Zhang, Z. Chen, JACS **130**, 16739 (2008).

[13] P. Murugan, V. Kumar, Y. Kawazoe and N. Ota, Phys. Rev. A **71**, 063203 (2005).

[14] A. Vojvodic, B. Hinnemann and J. K. Nørskov, Phys. Rev. B **80**, 125416 (2009).

[15] A. R. Botello-Mendez, F. Lopez-Urias, M. Terrones and H. Terrones, Nanotechnology, **20**, 325703, (2009).

[16] R. Shidpour and M. Manteghian, Nanoscale, **2**, 1429 (2010).

[17] P. Esquinazi, D. Spemann, R. Hohne, A. Setzer, K. H. Han and T. Butz, Phys. Rev. Lett. **91**, 227201 (2003).

[18] S. Mathew, B. Satpati, B. Joseph, B. N. Dev, R. Nirmala, S. K. Malik, R. Kesavamoorthy, Phy. Rev. B **75**, 075426 (2007).

[19] T. L. Makarova, A. L. Shelankov, I. T. Serenkov, V. I. Sakharov and D. W. Boukhvalov, Phys. Rev. B **83**, 085417 (2011).

[20] H. Ohldag, T. Tyliszczak, R. Hohne, D. Spemann, P. Esquinazi, M. Ungureanu, and T. Butz, Phys. Rev. Lett., **98**, 187204, (2007).

[21] X. J. Yu, O. Wilhelmi, H. O. Moser, S. V. Vidyarai, X. Y. Gao, A. T. S. Wee, T. Nyunt, H. Qian, and H. Zheng, J. Electron Spectrosc. Relat. Phenom. **144–147**, 1031 (2005).

[22] See Supplemental Material for experimental details and 0.5 MeV irradiation results.

[23] H. P. Myers, *Introductory Solid State Physics*, (Taylor and Francis, Oxford, 1997).

[24] H. Ohldag, P. Esquinazi, E. Arenholz, D. Spemann, M. Rothermel, and A. Setzer, New. J. Phys. **12**, 123012 (2010).

[25] J. R. Lince, T. B. Stewart, M. M. Hills, P. D. Fleischauer, J. A. Yarmoff, A. Taleb-Ibrahimi, Surf. Sci. **210**, 387 (1989).

[26] T. A. Patterson, J. D. Carver, D. E. Leyden, and D. M. Hercules, J. Phys. Chem. **80**, 1700 (1976).

[27] S. Mathew, T. K. Chan, D. Zhan, K. Gopinadhan, A. R. Barman, M. B. H. Breese, S. Dhar, Z. X. Shen, T. Venkatesan and John TL Thong, J. Appl. Phys. **110**, 84309 (2011).

[28] T. J. Wieting and J. L. Verble, Phys. Rev. B **3**, 4286 (1971).

[29]T. Sekine, K. Uchinokura, T. Nakashizu, E. Matsuura and R. Yoshizaki, J. Phys. Soc. Japan **53**, 811 (1984).

[30]G. L. Frey, R. Tenne, M. J. Matthews, M. S. Dresselhaus, and G. Dresselhaus, Phys. Rev. B **60**, 2883 (1999).

[31]N. Wakabayashi, H. G. Smith, and R. M. Nicklow, Phys. Rev. B **12**, 659 (1975).

[32]T. Livneh and E. Sterer, Phys. Rev. B **81**, 195209 (2010).

[33]R. K. Tiwari, J. Yang, M. Saeys, C. Joachim, Surf. Sci. **602**, 2628 (2008).

[34]Roger St. C. Smart, W.M. Skinner and A. R. Gerson, Surf. Interface Anal. **28**, 101–105 (1999).

[35]J. R. Lince, D. J. Carre, and P. D. Fleischauer, Langmuir **2**, 805 (1986).

[36]J. F. Ziegler, J. P. Biersack, U. Littmark, *The stopping and range of ions in matter* (Pergamon, New York, 1995).

[37]H. P. Komsa, J. Kotakoski, S. Kurasch, O. Lehtinen, U. Kaiser and A. V. Krasheninnikov, Phys. Rev. Lett. **109**, 035503 (2012).

[38]T. Venkatesan, R. Levi, T. C. Banwell, T. Tomberllo, M. Nicolet, R. Hamm and E. Mexixner, Mat. Res. Soc. Symp. Proc. **45**, 189 (1985).

[39]A. Benyagoub and A. Audren, J. Appl. Phys. **106**, 83516 (2009).

[40]P. O. Lehtinen, A. S. Foster, Y. Ma, A. V. Krasheninnikov, and R. M. Nieminen, Phys. Rev. Lett. **93**, 187202 (2004).

[41]L. Tsetseris and S. T. Pantelides, Phys. Rev. B **84**, 195202 (2011).

[42]Y. Zhang, S. Talapatra, S. Kar, R. Vajtai, S. K. Nayak and P. M. Ajayan, Phys. Rev. Lett. **99**, 107201 (2007).

# Figures

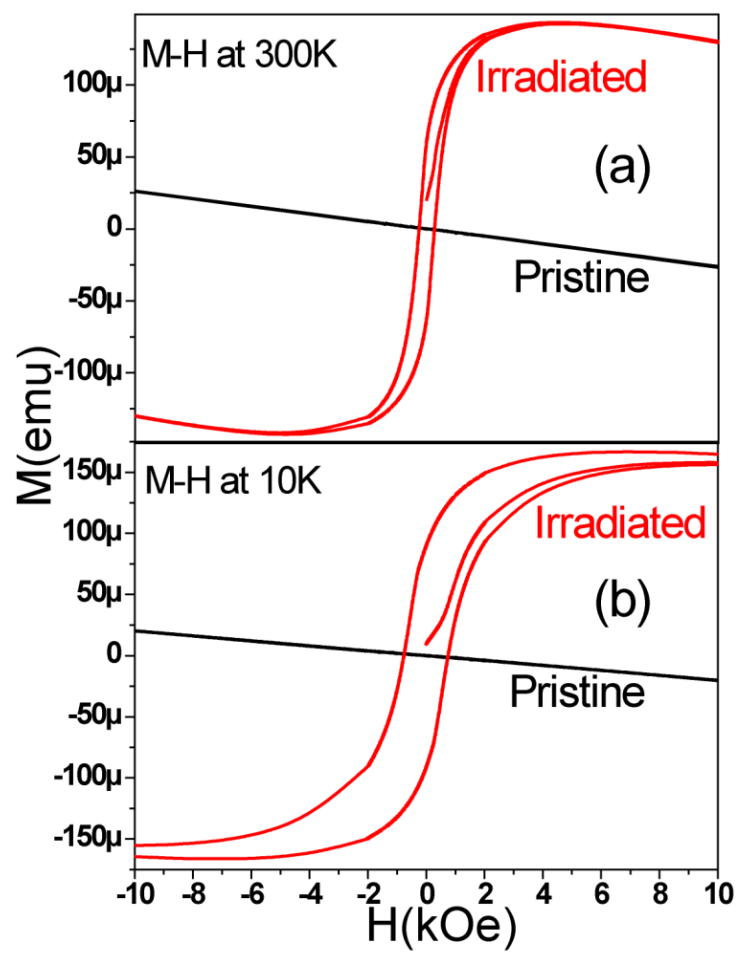

FIG. 1. *M* vs *H* curve (a) at 300 K and (b) at 10 K for a pristine and irradiated MoS$_2$ after subtracting the substrate Si contribution.

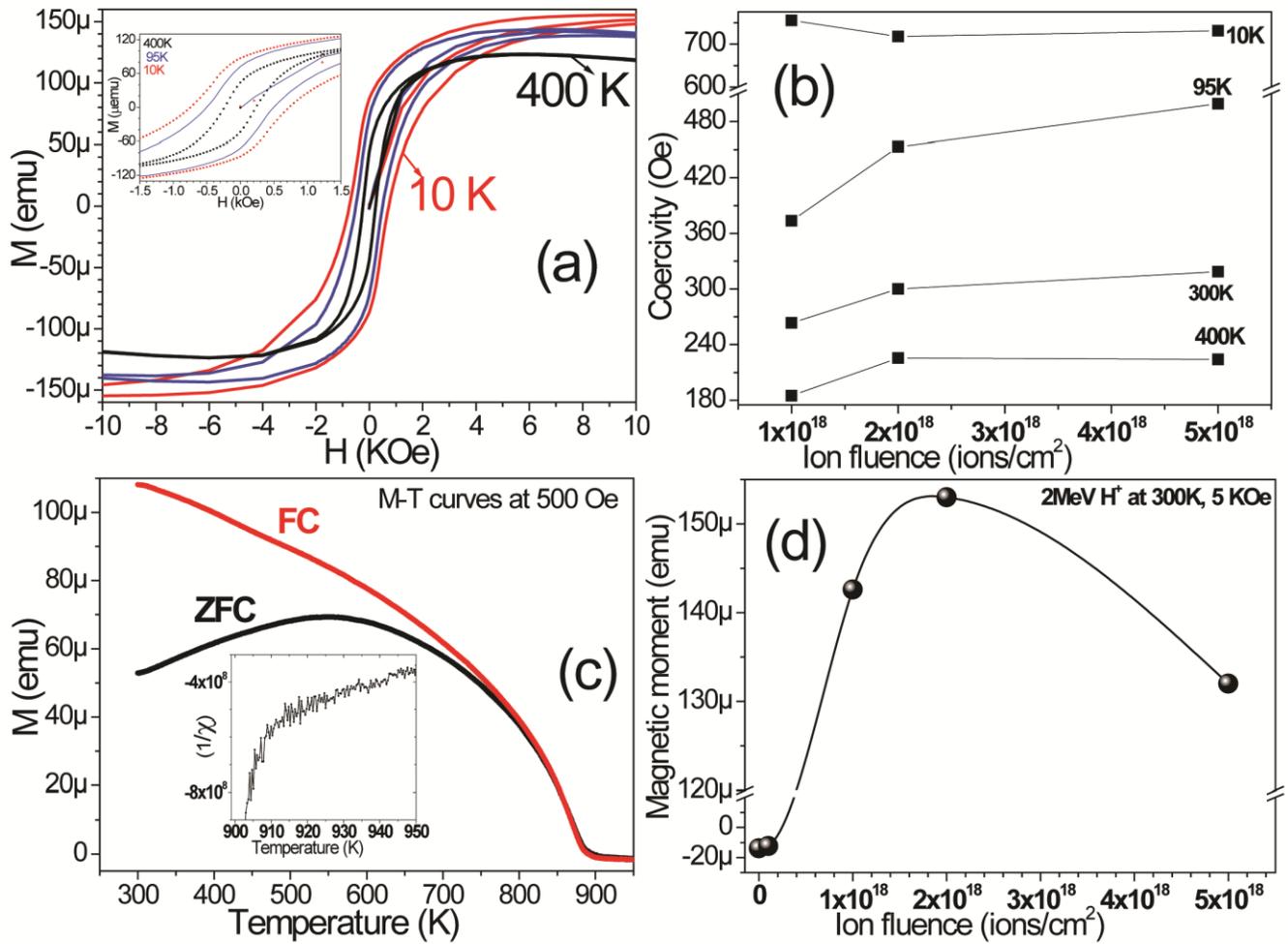

FIG. 2. (a)- *M* vs *H* curve at various temperatures from 400 K (black), 95 K (blue) and 10 K (red) for a pristine and irradiated $MoS_2$ at a fluence of $5 \times 10^{18}$ ions/cm$^2$. An enlarged view of the M-H isotherms near the origin is shown in the inset. The variation of coercivity *vs* ion fluence is shown in 2(b). 2(c)- Zero field and field cooled (ZFC and FC) magnetization *vs* temperature measurements in an applied field of 500 Oe for an irradiated $MoS_2$ sample at a fluence of $5 \times 10^{18}$ ions/cm$^2$. The inverse of the estimated magnetic susceptibility vs Temperature plot near $T_c$ (900-950 K) is shown in the inset. 2(d)- The value of magnetization at 300 K with H = 5 kOe as a function of ion fluence from $1 \times 10^{17}$ ions/cm$^2$ to $5 \times 10^{18}$ ions/cm$^2$.

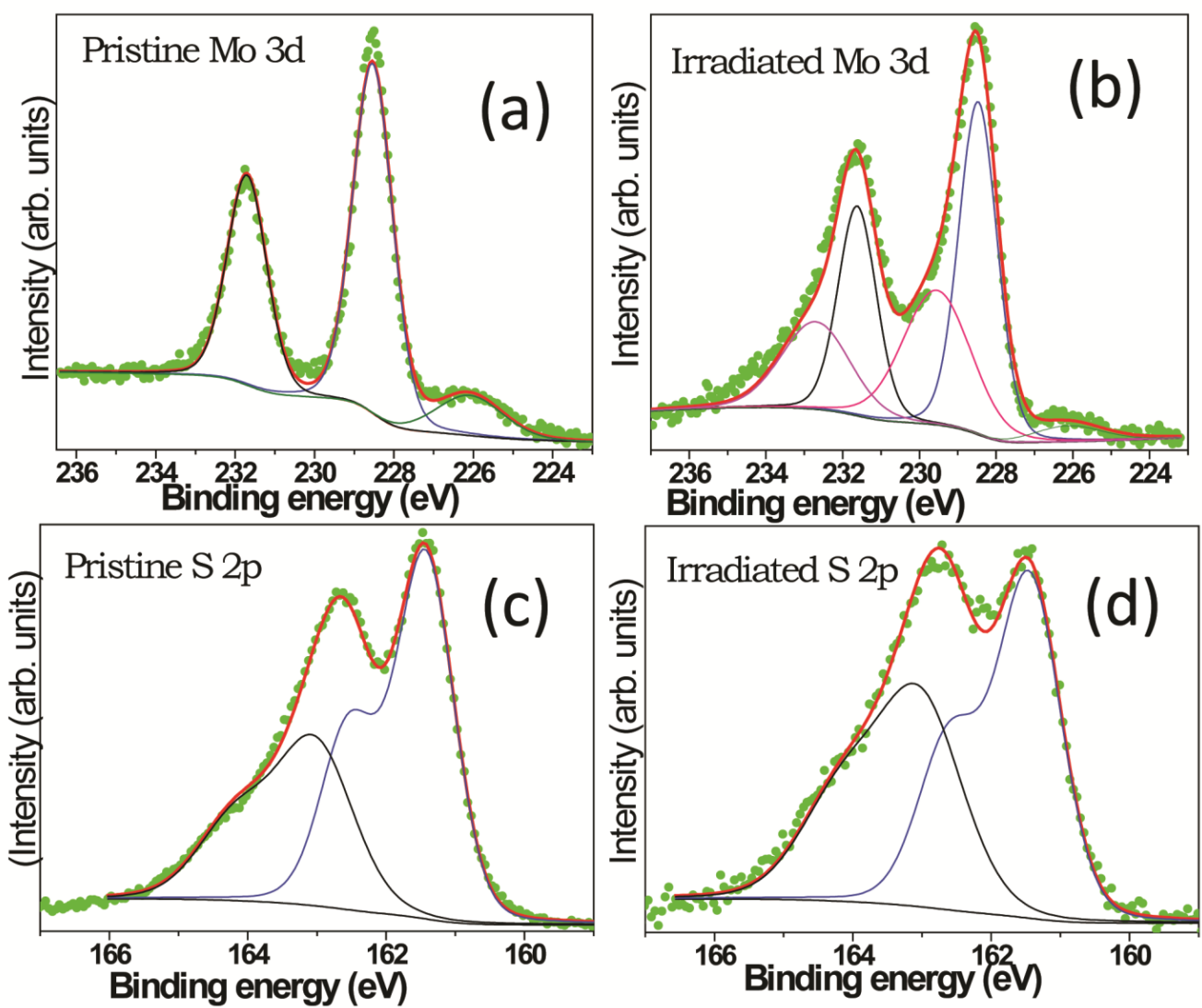

FIG. 3. XPS spectra from pristine and irradiated $MoS_2$ at a fluence of $5 \times 10^{18}$ ions/cm$^2$. The Mo peak is given in 3 (a) and (b) and S peak in 3(c) and (d). The fitted spectra along with the constituent peaks and experimental points are also shown.

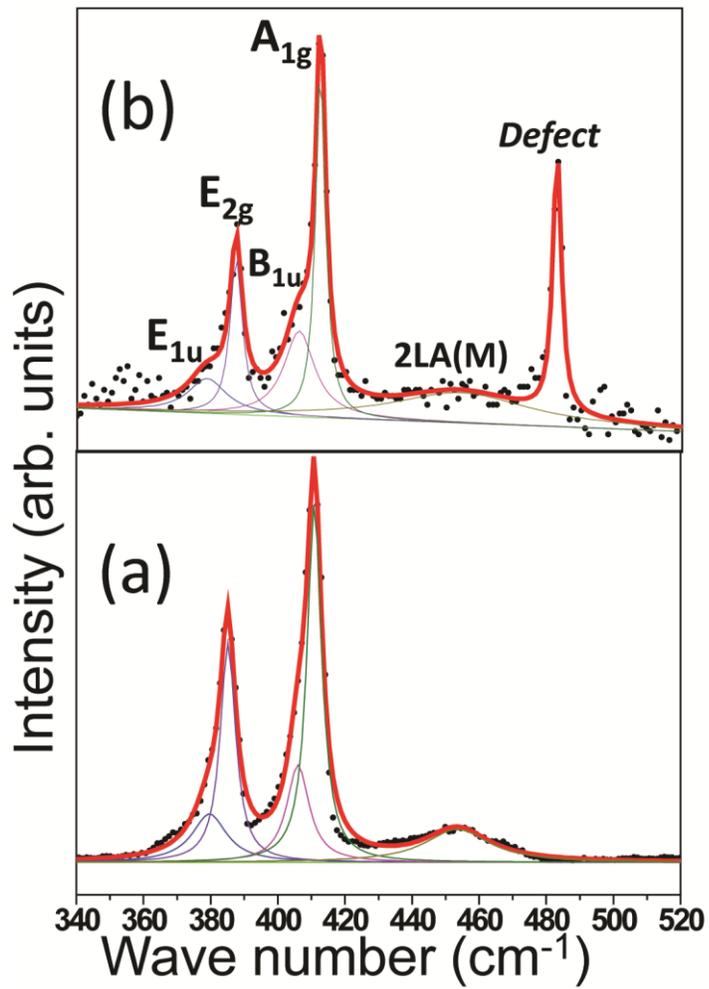

FIG. 4. Raman spectra of (a) pristine $MoS_2$ and (b) irradiated $MoS_2$ at a fluence of $5 \times 10^{18}$ ions/cm$^2$. The deconvoluted modes are labeled in the spectrum, the fitted curve with constituent peaks and experimental points are also given.